\documentclass[%
reprint,
superscriptaddress,
showpacs,
 amsfonts,amsmath,amssymb,
aip,
apl,
floatfix
]{revtex4-1}

\usepackage{graphicx}
\usepackage{dcolumn}
\usepackage{bm}

\begin{document}

\title{Electronic and optical properties of potential solar absorber Cu$_{3}$PSe$_{4}$}

\author{D.~H.~Foster}
\affiliation{Department of Physics, Oregon State University, Corvallis, Oregon 97331, USA}

\author{V.~Jieratum}
\author{R.~Kykyneshi}
\author{D.~A.~Keszler}
\affiliation{Department of Chemistry, Oregon State University, Corvallis, Oregon 97331, USA}

\author{G.~Schneider}
\email{Guenter.Schneider@physics.oregonstate.edu}
\affiliation{Department of Physics, Oregon State University, Corvallis, Oregon 97331, USA}

\date{\today}

\begin{abstract}
We theoretically investigate the electronic and optical properties of semiconductor Cu$_{3}$PSe$_{4}$.
We also report diffuse reflectance spectroscopy measurements for Cu$_{3}$PSe$_{4}$~and Cu$_3$PS$_4$, which indicate a band gap of 1.40 eV for the former.
Hybrid functional calculations agree well with this value, and reveal that the band gap is direct.
Calculations yield an optical absorption spectrum similar to GaAs in the visible region, with $\alpha > 5 \times 10^4$ cm$^{-1}$ for $\lambda < 630$~nm.
We conclude that the optical properties of Cu$_{3}$PSe$_{4}$~are within the desired range for a photovoltaic solar absorber material.
\end{abstract}

\maketitle

The need for low cost solar photovoltaic cells continues to drive a global effort to find, evaluate, and refine materials that offer both cost effective and highly efficient solutions.
Within the search among the inorganic semiconductors, one of the most exciting families has been the ternary and quaternary copper chalcogenides consisting of CuCh$_4$ and ACh$_4$ tetrahedra, where Ch is one of (S, Se, Te) and A represents one or more other elements.
Examples include the commercially widespread Cu(In,Ga)Se$_2$ (CIGS), as well as Cu$_2$ZnSnS$_4$, which has the benefit of consisting of low-toxicity, readily available elements.
One material in this family which has not received much attention in this regard is Cu$_{3}$PSe$_{4}$.
The structure of Cu$_{3}$PSe$_{4}$~has been determined\cite{garin_crystal_1972,ma_synthesis_2002}, and a photoelectrochemical analysis\cite{marzik_photoelectronic_1983} of isostructural compounds Cu$_3$PS$_4$ and Cu$_3$PS$_3$Se has indicated indirect bandgaps of 2.38 eV and 2.06 eV respectively.
However the bandgap of Cu$_{3}$PSe$_{4}$~itself has not been reported previously.
In this Letter we examine this material, theoretically in the scope of low temperature and zero defects, and experimentally in the scope of room temperature measurements on powder samples.

\begin{figure}[b]
\includegraphics{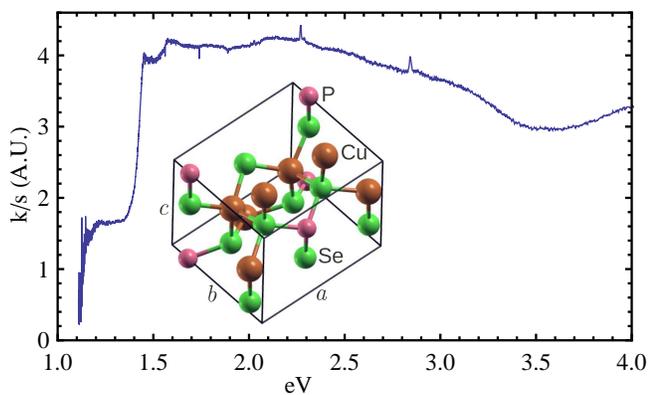}
\caption{\label{fig_exp_and_struct}(Color online) Measured diffuse reflectance of Cu$_{3}$PSe$_{4}$. Inset: The wurtzite-based enargite structure. For Cu$_{3}$PSe$_{4}$, $a=7.685$ \AA, $b=6.656$ \AA, $c=6.377$ \AA.}
\end{figure}

In our synthesis of Cu$_{3}$PSe$_{4}$~and Cu$_3$PS$_4$, elemental powders of Cu (Cerac, 99.5\%), P (Alpha Aesar 99\%), Se (Cerac, 99.6\%), and S (Cerac, 99.999\%), are ground and sealed in evacuated fused silica tubes.
Polycrystalline powders are obtained via solid state reaction at 480 $^\circ$C (Cu$_{3}$PSe$_{4}$) and 600 $^\circ$C (Cu$_3$PS$_4$) for 24 h.
The XRD patterns, collected on a Rigaku Ultima IV, of the reacted samples are similar to the ISCD card $\#$095412 for Cu$_{3}$PSe$_{4}$~and $\#$412240 for Cu$_3$PS$_4$.
Hole majority carrier type is confirmed by positive thermoelectric voltage measurements.
The optical bandgaps are determined from diffuse reflectance measurements (see Figure \ref{fig_exp_and_struct}) collected with a spectrometer equipped with an Ocean Optics HR4000 UV-VIS detector and a balanced deuterium/tungsten halogen source (DH-2000-BAL).
The Kubelka-Munk model is employed to determine the optical absorption edges of 1.40 eV and 2.38 eV for Cu$_{3}$PSe$_{4}$~and Cu$_3$PS$_4$, respectively.
The Cu$_{3}$PSe$_{4}$~bandgap is within the desired range for solar photovoltaic devices (1.0 eV to 1.6 eV), while the Cu$_3$PS$_4$ bandgap agrees precisely with Ref.~[\onlinecite{marzik_photoelectronic_1983}].

Calculations based on density functional theory (DFT) are performed using the projector augmented wave (PAW) method\cite{bloechl_projector_1994,*kresse_ultrasoft_1999} as implemented in the plane wave code \texttt{VASP}\cite{kresse_efficient_1996}.
For the exchange-correlation functional we use the generalized gradient approximation (GGA) in the PW91 parametrization\cite{perdew_atoms_1992} for accurate total energy calculations.
To avoid the bandgap underestimation common to standard DFT calculations\cite{stampfl99den}, we use the Heyd-Scuseria-Ernzerhof\cite{heyd_hybrid_2003,*heyd_erratum:_2006} (HSE) hybrid functional, which combines Hartree-Fock (HF) exchange with GGA exchange and includes an empirical shielding of the HF exchange.
The HSE functional has yielded respectable estimates for semiconductor bandgaps\cite{peralta_spin-orbit_2006,*kim_accurate_2009,*chan_efficient_2010}, although it is generally not as reliable as methods based on the $GW$ approximation.\cite{shishkin2007sel}
We will see below that HSE performs particularly well for calculations on Cu$_{3}$PSe$_{4}$.
For brevity and the expressed application interest, we present theoretical analysis for Cu$_{3}$PSe$_{4}$~but not Cu$_3$PS$_4$.
Both materials crystallize in the wurtzite-based enargite structure with a simple orthorhombic unit cell with 16 atoms admitting the space group $Pmn2_1$ (FIG.~\ref{fig_exp_and_struct}). 
For electronic and optical calculations, we use experimentally determined lattice and atomic parameters.\cite{ma_synthesis_2002}
Ionic and lattice relaxations are discussed below.
In all calculations, we carefully consider the convergence of the result with respect to plane wave basis cut-off energy, $k$-point density, and time saving approximations.

\begin{figure}[t]
\includegraphics{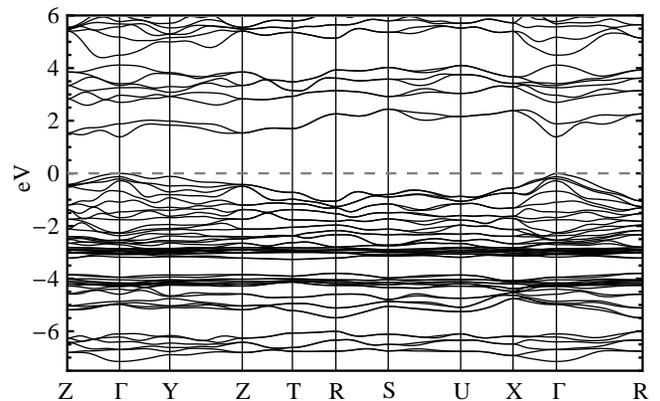}
\caption{\label{fig_bands}Band structure of Cu$_{3}$PSe$_{4}$~determined by HSE calculation. The direct bandgap of 1.38 eV lies at $\Gamma$.}
\end{figure}

The HSE determined band structure is shown in FIG.~\ref{fig_bands}.
The HSE calculations used a $\Gamma$-centered $8 \times 8 \times 8$ $k$-point grid with $2\times$ down-sampling for one of the $k$-point summations in the evaluation of the Hartree-Fock exchange potential.\cite{paier_screened_2006}
The band structure plot itself (FIG.~\ref{fig_bands}) is formed by interpolation of the $8^3$ $k$-point grid using a Fourier method.\cite{pickett88smo}
The band structure most importantly shows a direct bandgap of 1.38 eV at the $\Gamma$ point, in excellent agreement with the experimental value of $1.40$ eV.
The band structure also shows low lying valleys ($\sim0.1$ eV) in both the valence band (at Y) and the conduction band (along $\Gamma$-Z).
We note that an HSE calculation for Cu$_3$PS$_4$ at experimental parameters\cite{pfitzner_refinement_2002} gives these two valleys as being the valence band maximum (VBM) and conduction band minimum (CBM), with theoretical and experimental\cite{marzik_photoelectronic_1983} indirect bandgaps of 2.55 eV and 2.38 eV respectively.
We calculate electron and hole effective masses of 0.17 and 0.60 times the electron mass, respectively.
Spin-orbit coupling was verified to be small in a single GGA calculation which showed VBM splitting and a bandgap decrease of 0.030 eV relative to the spinless GGA bandgap of 0.29 eV.

An analysis of the partial density of states indicates that the valence bands from the VBM to about $-5.5$ eV below (see FIG.~\ref{fig_bands}) have similar orbital composition to the corresponding bands of Cu$_2$ZnSnS$_4$.\cite{paier09cu2}
In particular, the valence band edge has anti-bonding $t_{2}$ Cu-d/Se-p$^*$ hybrid character, with the corresponding bonding states lying in the second valence band.
Between these regions, at the bottom of the first valence band, lies the very dense bundle of non-bonding $e$ Cu-d orbitals.
The two states in the lowest conduction band show hybridization among Se-p, P-s, and Cu-d orbitals.
The charge density distributions of these states indicate significant anti-bonding character between P and Se ions.
This will have important effects on relaxation calculations, as discussed below.

\begin{figure}
\includegraphics{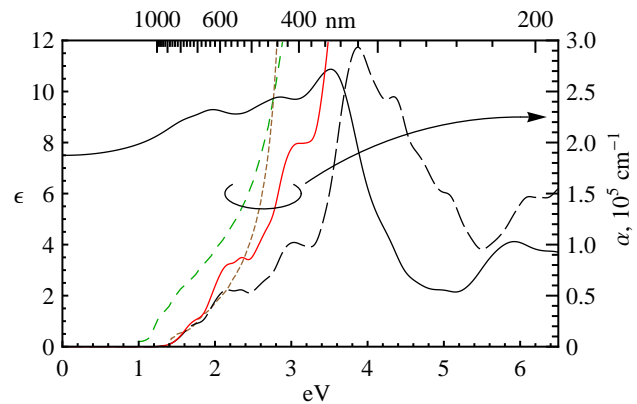}
\caption{\label{fig_abseps}(Color online) Real (solid) and imaginary (long-dashed) parts of dielectric function $\epsilon$ averaged over principal axes.
The absorption coefficient $\alpha$ is shown (scale on right) for Cu$_{3}$PSe$_{4}$~(red solid), GaAs (brown short-dashed), and CIGS (green dashed).}
\end{figure}

To obtain the optical properties of Cu$_{3}$PSe$_{4}$, the complex dielectric tensor $\epsilon$ has been calculated\cite{gajdos_linear_2006} in the random phase approximation from HSE wavefunctions without including local effects, which are generally small.\cite{footnote1}
The direction-averaged dielectric function $\text{Tr}(\epsilon) / 3$ and optical absorption coefficient $\alpha$ are shown in Figure~\ref{fig_abseps}.
Optical absorption is compared with experimental data\cite{palik_handbook_1991,paulson_optical_2003} for GaAs and polycrystalline CuIn$_{1-x}$Ga$_{x}$Se$_2$ for $x=0.31$.
(In photovoltaic applications, CIGS stoichiometries near $x = 0.3$ yield the most efficient power conversion.\cite{kemell05thi})
For $520 < \lambda < 660$~nm, the absorption in Cu$_{3}$PSe$_{4}$~is calculated to be higher than in GaAs.
The greater absorption of CIGS is primarily due to its absorption curve being shifted to lower energies by its smaller bandgap (1.17 eV).
While allowing absorption of a greater fraction of the solar spectrum, the CIGS bandgap limits its solar cell voltage---and its maximum theoretical efficiency---relative to larger bandgap materials.
Absorption is above $5\times 10^4$ cm$^{-1}$ for wavelengths $\lambda < 630$~nm ($\hbar \omega > 2.0$ eV)\cite{[{This data point is approximately shared with Cu$_2$ZnSnS$_4$ which has similar absorption. }]ito88ele} and above $1 \times 10^5$ cm$^{-1}$ for $\lambda < 480$~nm ($\hbar \omega > 2.6$ eV).

As discussed previously, the HSE bandgap using the experimental atomic and lattice parameters agrees well with the measured bandgap.
However, using GGA or local density approximation (LDA) relaxed structural parameters, we calculate HSE bandgaps approximately 30\% smaller (TAB. \ref{tab:relax}).
While the bandgap is expected to depend on the lattice parameters, such a large change is unusual, but can readily be understood as follows: An HSE ion relaxation leaves the experimental structure and bandgap nearly unchanged. The LDA and GGA relaxations, however, result in a lengthening of the P-Se bond, accompanied by large reductions of the HSE bandgap.
The observed relation between the increase of the P-Se bond length and the underestimate of the calculated bandgaps is a direct consequence of the significant P-s/Se-p$^*$ anti-bonding character of the states at the conduction band edge: an underestimate of the bandgap (as in LDA/GGA) moves the P-s/Se-p$^*$ anti-bonding orbitals to lower energies, resulting in a smaller P-Se bond energy and a correspondingly longer bond.
A similar bond length dependence of the bandgap of CuInSe$_2$ is discussed in Ref.~[\onlinecite{jaffe1984the}].

\begin{table}
\caption{\label{tab:relax}Comparison of bandgap and structural quantities for different exchange-correlation functionals.
Both ion-only and full ion+lattice relaxations are performed.
All bandgaps are predicted to be direct and at $\Gamma$ unless noted otherwise.
$E_{\text{g}}$ (HSE) is the HSE-calculated direct bandgap. 
$E_{\text{g}}$ ($X$) denotes the direct bandgap given by measurement or by the functional used in the relaxation (LDA, GGA, or HSE).
$V$ denotes cell volume and the bond lengths are averages.
LDA and GGA calculations using the unrelaxed experimental structure result in bandgaps of 0.21 and 0.29 eV, respectively.}
\begin{ruledtabular}
\begin{tabular}{llllll}
relaxation & $E_{\text{g}}$ (HSE) & $E_{\text{g}}$ ($X$) & $V$ & $\langle$P-Se$\rangle$ & $\langle$Cu-Se$\rangle$ \\
& (eV) & (eV) & (\AA$^3$) & (\AA) & (\AA) \\ 
\hline\\[-1.5ex]
LDA ion & 0.94 & 0.08\footnotemark[1] & 326.19 & 2.30 & 2.39 \\
LDA full & 1.10\footnotemark[2] & 0.06 & 310.59 & 2.27 & 2.35 \\
GGA ion & 1.02 & 0.03 & 326.19 & 2.29 & 2.40 \\
GGA full & 0.90 & 0.08\footnotemark[1] & 338.25 & 2.31 & 2.43 \\
HSE ion & 1.42 & 1.42 & 326.19 & 2.23 & 2.42 \\
exp.\cite{ma_synthesis_2002} & 1.38 & 1.40 & 326.19 & 2.24 & 2.41 \\
\end{tabular}
\end{ruledtabular}
\footnotetext[1]{An indirect bandgap is predicted (0.04 eV, VBM along $\Gamma$-Y in both cases).}
\footnotetext[2]{An indirect bandgap is predicted (1.04 eV, CBM along $\Gamma$-Z).}
\end{table}

A further structural analysis examines possible alternatives to the enargite structure used in the preceding calculations.
While only the enargite form of Cu$_{3}$PSe$_{4}$~has been observed experimentally, it is possible that competing structural phases could form under synthesis techniques appropriate for device construction.
To examine structural stability, we have performed a set of GGA calculations (Table \ref{tab:hof}) comparing the heat of formation $\Delta H$ for Cu$_{3}$PSe$_{4}$~placed in several structures which are manifested by other compounds having the form A$_3^{+}$B$^{5+}$C$_4^{2-}$.
The $\Delta H$ values of the famatinite\cite{pfitzner_refinement_2002} and enargite structures differ negligibly.
This is not surprising, as the essential difference of the structures lies only in the arrangement of the (Cu,P)Se$_4$ tetrahedra: zincblende for famatinite and wurtzite for enargite.
Using the lattice and atomic parameters obtained from fully relaxed GGA calculations, we have calculated HSE bandgaps of 0.90 eV (enargite) and 0.84 eV (famatinite) for these potentially competing structures.
Though artificially low due to the GGA relaxations, the similarity of these bandgaps indicates that the presence of a famatinite phase would not greatly alter the optical properties of Cu$_{3}$PSe$_{4}$.
\begin{table}
\caption{\label{tab:hof}Calculated zero temperature heat of formation ($\Delta H$) 
for Cu$_{3}$PSe$_{4}$~in several different imaginable structures. 
$\Delta H$
 is relative to fully relaxed elements (allotropes: black P and $\gamma$-Se) and is given per 8-atom formulaic unit. Volume, cell shape, and ion positions are relaxed at fixed symmetry.} 
\begin{ruledtabular}
\begin{tabular}{llD{.}{.}{-1}}
structure & space group & \multicolumn{1}{c}{$\Delta H$ (eV)} \\ \hline
enargite (Cu$_3$PS$_4$) & $Pmn2_1$ (no.~31) & -1.324 \\   
famatinite (Cu$_3$SbS$_4$) & $I\bar{4}2m$ (no.~121) & -1.309 \\
lazarevicite (Cu$_3$AsS$_4$) & $P\bar{4}3m$ (no.~215) & -1.106 \\
Rb$_3$PS$_4$ & $Pnma$ (no.~62) & -0.500 \\
Na$_3$PS$_4$ & $P\bar{4}2_1c$ (no.~114) & 0.107
\end{tabular}
\end{ruledtabular}
\end{table}

In conclusion, the measurements and calculations reported here indicate that Cu$_{3}$PSe$_{4}$~has optical properties which make it viable for photovoltaic applications.
Further measurement and calculation of its properties, thermodynamic stability, and primary defects will be subsequently reported, and will further analyze the potential of Cu$_{3}$PSe$_{4}$~as a photovoltaic or photoelectronic material.
This work has been supported by National Science Foundation grant SOLAR DMS-1035513.


\begin{thebibliography}{25}
\bibitem {footnote1}%
  \BibitemOpen
  \bibinfo {note} {For optical properties, the full HSE calculation was
  repeated using a finer $12^3$ $k$-point grid with $3\times $ down-sampling.
  We note that optical properties calculated with the ``ground state'' HSE
  method used here have been shown in Ref.~[\protect \rev@citealpnum
  {paier09cu2}] to be very similar to results of the more accurate time
  dependent HSE method.}\BibitemShut {Stop}%
\end{thebibliography}
\end{document}